\documentclass[12pt]{iopart}
\usepackage{graphicx}
\newcommand{\beq}{\begin {equation}}
\newcommand{\eeq}{\end {equation}}

\def\hpp{{\hat{\bf p}}'}
\def\hqp{{\hat{\bf q}}'}

\begin{document}

\title[]
{Faddeev calculation for breakup neutron-deuteron scattering at 
14.1 MeV lab energy}
\author{V M Suslov$^1$, I Filikhin$^1$ and B Vlahovic$^1$,}
\address{$^1$Physics Department, North Carolina Central University,
       1801 Fayetteville Street, Durham, NC 27707, USA}  
\author{M A Braun$^{1,2}$,} 
\address{$^2$Department of Theoretical Physics,
        Saint-Petersburg State University,
        198504 Ul'yanovskaya 1, Petrodvorets, Saint-Petersburg, Russia}  
\author{I Slaus$^{1,3}$}
\address{$^3$R. Boskovic Institute, 10000 Zagreb, Croatia}
 
\ead{vsuslov@nccu.edu}

\begin{abstract}
A new computational method for solving the nucleon-deuteron breakup scattering problem has been applied to study the inelastic neutron-deuteron scattering on the basis of the configuration-space Faddeev equations. This method is based on the spline-decomposition in the angular variable and on a generalization of the Numerov method for the hyperradius. The Merkuriev-Gignoux-Laverne approach has been generalized for arbitrary nucleon-nucleon potentials and with an arbitrary number of partial waves. Neutron-deuteron observables at the incident nucleon energy 14.1 MeV have been calculated using the charge-independent AV14 nucleon-nucleon potential. Results have been compared with those of other authors and with experimental neutron-deuteron scattering data.
\end{abstract}
\maketitle

\section{Introduction}
The last two decades brought tremendous progress in both
theoretical and experimental study of nucleon-deuteron scattering.
More accurate experimental data were compared with rigorous
calculations in the framework of the Faddeev equations with
high-precision nucleon-nucleon potentials also including model
three-nucleon forces \cite{Gloec,Delt1}. Nucleon-nucleon (NN)
potentials used in rigorous three-nucleon (3N) calculations are
charge-independent AV14 \cite{Wir1}, AV18 \cite{Wir2}, CD-Bonn
\cite{Mach1} and several Nijmegen potentials \cite{Stoc1}. Among
three-nucleon forces (3NF) are Tucson-Melbourne (TM) and its
various modifications \cite{Coon}, and Urbana potentials
\cite{Pudi}. Based on the chiral effective field theory (EFT) NN
and 3N potentials have been constructed \cite{Mach2} and  used in
rigorous 3N calculations \cite{Delt1}.

     In spite of these enormous achievements  in the 3N studies,
there are several important cases where the rigorous 3N
calculations have failed to explain the data \cite{Ivo} and these
discrepancies are established with very high precision. Among the
most important discrepancies are the $\rm A_y$ puzzle in
nucleon-deuteron (Nd) elastic scattering \cite{Torn}, the star
configuration in the Nd breakup reaction \cite{Wiel} and the
quasi-free scattering (QFS) cross section \cite{Gloec}.

Recent calculations of the proton-deuteron breakup cross-section,
and tensor analyzing power data for a symmetric constant relative
energy (SCRE) geometry reported in \cite{Koeln} revealed a serious
disagreement between the theory and  experimental data. Inclusion
of the TM or Urbana 3NF did not solve the problem. The chiral EFT
showed the same deviations from the data. Including the Coulomb
force partially improved the  situation for the cross section. On
the other hand predictions in \cite{Maeda} with the 3NF forces and
Coulomb interaction taken into account showed a good agreement
with the data. At the same time, the recent study of Nd breakup in
symmetric star configurations based on Faddeev calculations with
CD-Bonn NN potential \cite {CRH} showed significant discrepancy
for nn QFS configuration and a good agreement with cross-section
data in the case of pn QFS breakup geometry. Thus the situation
with theoretical predictions for the 3N observables remains still
far from completely resolved. This motivates us to continue
searching for new computational procedures and input potentials,
which may allow to overcome the mentioned disagreements.

In this article we present  an alternative method for the study of
the  neutron-deuteron (nd) system based on the direct numerical
solution of the Faddeev equations in configuration space. This
approach was initiated by Merkuriev et al. (MGL) \cite{MGL} who
derived general formulae for nd breakup scattering. This method
has been initially applied to the study of nd and pd elastic and
breakup scattering with nuclear  interaction limited only to
S-waves and with simple NN potentials \cite{KKM}. In the present
work we generalize the MGL approach to any high precision
realistic potentials for nd inelastic processes.

This paper is organized as follows: in section 2 we describe
calculations in configuration space starting with the general
formalism in subsection 2.1 followed by Numerov method in
subsection 2.2. Our novel method for solution is given in
subsection 2.3. Formulae for calculating elastic and breakup
amplitudes in MGL basis are presented in section 2.4. Comparison
of our results with the previous calculations and with the data
are discussed in section 3. Finally, discussion and conclusion are
given in sections 4 and 5.
\section {Three-nucleon Faddeev calculation in configuration space - our new computational method}
\subsection {Formalism}
The starting point for studying interactions between nucleons in
three-body systems is the solution of the Schr\"odinger equation
$H\Psi = E\Psi$ for nuclear Hamiltonian
 \beq H =
-\frac{\hbar^2}{2m}\sum_{i=1}^3\nabla^2_i + \sum_{j<k} V_{jk}\ \
\Big(+\sum_{j<k<l} V_{jkl} \Big), \eeq where $V_{jk}$ and
$V_{jkl}$ are the two- and three-nuclear potentials, respectively.
In this study we neglected three-nucleon forces $V_{jkl}$.

     Writing the total wave function as
\beq
\Psi= \Phi_1+\Phi_2+\Phi_3 = (1+P^++P^-)\Phi_1,
\eeq
the Schr\"odinger equation for three identical particles can be reduced
into a single Faddeev equation, which in Jacobi's vectors $\vec x_1,\vec y_1$
has the form
\beq
\label{FNE}
\Big[-\frac{\hbar^2}{m}\Big(\Delta_{\vec x_1}+\Delta_{\vec y_1}\Big) +
V(\vec x_1) - E\Big]\Phi(\vec x_1,\vec y_1)=-V(\vec x_1)(P^++P^-)\Phi(\vec x_1,\vec y_1),
\eeq
where the operators $P^{\pm}$ are the cyclic permutation operators for
the three particles which interchange any pair of nucleons ($ P^+:
123 \rightarrow 231, P^-: 123 \rightarrow 312$), and $\hbar ^2/m$=41.47 MeV$\cdot\mbox
{fm}^2$. As independent coordinates, we take the Jacobi vectors ${{\vec x}_\alpha , {\vec y}_\alpha }$. For the pair $\alpha $=1, they are related to particle coordinates by the formulas:
\begin{equation}
{\vec x}_1={\vec r}_2-{\vec r}_3, \ \ \ \ \ \ {\vec y}_1 =%
\frac{{\vec r}_2+{\vec r}_3}2 -{\vec r}_1,
\end{equation}
for $\alpha $=2,3 one has to make cyclic permutations of the indexes in Eq.(5). The Jacobi vectors with
different $\alpha$'s are linearly related by the orthogonal transformation
\begin{equation}
  \left(
  \begin{array}{c}
     {\vec x}_{\alpha} \\ {\vec y}_{\alpha}
  \end{array}
  \right)=
  \left(
  \begin{array}{rl}
      C_{\alpha\beta} & S_{\alpha\beta} \\
     -S_{\alpha\beta} & C_{\alpha\beta}
  \end{array}
  \right)
  \left(
  \begin{array}{c}
     {\vec x}_{\beta} \\ {\vec y}_{\beta}
  \end{array}
  \right) \ ,\ \ \ C^2_{\alpha\beta} + S^2_{\alpha\beta} = 1,
\end{equation}
where
\begin{equation}
\begin{array}{c}
C_{\alpha\beta}=-\sqrt{\frac{m_{\alpha}m_{\beta}}
{(M-m_{\alpha})(M-m_{\beta})}}, \ \
S_{\alpha\beta} = (-)^{\beta - \alpha}{\rm sgn}(\beta - \alpha)
\sqrt{1-C^{2}_{\alpha\beta}},  \\  M=\sum_{\alpha=1}^3m_{\alpha}.
\end{array}
\end{equation}

To perform numerical calculations for arbitrary nuclear potential, we use MGL approach \cite{MGL}. To make the angular analysis of Eq.(\ref{FNE}), we use a basis proposed in \cite{MGL}. This basis is intermediate between LS and Jj coupling schemes.\\
Let: $\vec{\sigma}$,  $\bf l$ and ${\bf J}=\vec{\sigma}+{\bf l}$ be the spin, orbital and total angular momenta of the pair 23. ${\bf s=1/2+J}$ the total "spin" of the system 123 considering the pair 23 as a particle with "spin" $\bf J$. $\vec{\lambda}$ the orbital momentum conjugate to $\bf y$, that is of the relative motion of particle 1 respective to the c.m of pair 23. $\bf {M}=\vec{\lambda}+\bf{s}$ the total angular
momentum with its $z$-projection $M_z$. To this we have to add the isospin part. If the isospin of the pair 23 is $\bf t$, then the total isospin is ${\bf T=1/2+t}$ with its projection $T_z$. Since in the nd case $T=1/2$ it need not be shown explicitly. The set of quantum numbers $\{\lambda s l\sigma J\}\equiv\alpha$ defines a state in this basis.

Correspondingly in this basis the spin-angular-isospin eigenfunctions have the form
\beq
 Z_{\alpha}({\bf
\hat{x},\hat{y}})=<{\bf \hat{x},\hat{y}}|\alpha>=
\Big[Y^{\lambda}({\bf \hat{y}})\otimes[\chi^{1/2}\otimes{\cal
Y}^J_{l\sigma} ({\bf\hat{x}})]^s\Big]^{M,M_z}\eta^{T,T_z}_{1/2,t}
\eeq

     For $nd$ scattering the Faddeev equations for partial components can be written in the following form (here we omit the index 1):
\begin{equation}
\label{fexy}
\begin{array}{c}
\Big[E+\frac{\hbar^2}{m}(\partial_x^2+\partial_y^2)-v_{\alpha}^{\lambda l}(x,y)
\Big]\Phi^{\lambda_0,s_0,M_0}_{\alpha}(x,y) =
\\
\displaystyle
\sum_{\beta}v_{\alpha\beta}(x)\Big[\Phi^{\lambda_0,s_0,M_0}_{\beta}(x,y)
+\int_{-1}^1du\sum_{\gamma}g_{\beta\gamma}(y/x,u)
\Phi^{\lambda_0,s_0,M_0}_{\gamma}(x',y')\Big].
\end{array}
\end{equation}

     The geometrical function $g_{\beta\gamma}(x,y,u)$ is the representative of the permutation operator $P^++P^-$
in MGL basis \cite{MGL}:
\[\
g_{\alpha'\alpha}(y/x,u)=g_{\alpha'\alpha}(\theta,u)=
g_{\alpha'\alpha}(\theta,\theta')\]\[=
(-1)^{\lambda+\lambda'+J+J'}
[(2J+1)(2J'+1)(2s+1)(2s'+1)]^{1/2}
\sum_{LS}(2S+1)(2L+1)\]\[
    \left \{ \begin{array}{ccc}
    l     & \sigma & J \\
    1/2 & s & S
    \end{array} \right \}
 \left \{ \begin{array}{ccc}
    l'     & \sigma' & J' \\
    1/2 & s' & S
    \end{array} \right \}
 \left \{ \begin{array}{ccc}
    \lambda     & l & L \\
    S & M & s
    \end{array} \right \}
 \left \{ \begin{array}{ccc}
    \lambda'     & l' & L \\
    S & M & s'
    \end{array} \right \}\]\beq
<\chi^S_{1/2\sigma'}\eta^T_{1/2,t'}|P^+|\chi^S_{1/2\sigma}\eta^T_{1/2,t}>
h^L_{\lambda'l'\lambda l}(y/x,u).
\eeq

Function $h$ is the representative of the permutation operator
$P^++P^-$ in the $\lambda+ l=L$ basis:
\[ h^L_{\lambda'l'\lambda l}(y/x,u)=h^L_{\lambda'l'\lambda
l}(\theta,u)= h^L_{\lambda'l'\lambda l}(\theta,\theta')\]\[=
\frac{xy}{x'y'}(-1)^{l+L}\frac{(2\lambda+1)(2l+1)}{2^{\lambda+l}}
[(2\lambda)!(2l)!(2\lambda'+1)(2l'+1)]^{1/2}\]\[
\sum_{k=0}(-1)^k(2k+1)P_k(u)\sum_{\lambda_1+\lambda_2=\lambda,\,
l_1+l_2=l}\frac{y^{\lambda_1+l_1}x^{\lambda_2+l_2}}{{y'}^{\lambda}{x'}^l}
(-1)^{l_2}\]\[\frac{(\sqrt{3})^{\lambda_2+l_1}}
{[(2\lambda_1)!(2\lambda_2)!(2l_1)!(2l_2)!]^{1/2}}
\sum_{\lambda''l''}(2\lambda''+1)(2l''+1) \]\[\left (
\begin{array}{ccc}
    \lambda_1     & l_1 & \lambda''\\
    0 & 0 & 0
    \end{array} \right )
\left ( \begin{array}{ccc}
    \lambda_2     & l_2 & l''\\
    0 & 0 & 0
    \end{array} \right )
\left ( \begin{array}{ccc}
    k     & \lambda'' & \lambda'\\
    0 & 0 & 0
    \end{array} \right )
\left ( \begin{array}{ccc}
    k     & l'' & l'\\
    0 & 0 & 0
    \end{array} \right )\]\beq
\left \{ \begin{array}{ccc}
    l'     & \lambda' & L\\
    \lambda'' & l'' & k
    \end{array} \right \}
\left \{ \begin{array}{ccc}
    \lambda_1     & \lambda_2 & \lambda\\
    l_1 & l_2 & l\\
    \lambda''&l''&L
    \end{array} \right \}.
\eeq
The index $k$ runs from zero to $(\lambda'+l'+\lambda+l)/2$.
The $(...)$ are the 3$j$ symbols:
\[
\left ( \begin{array}{ccc}
    j_1     & j_2 & j_3\\
    m_1 & m_2 & m_3
    \end{array} \right )= (-1)^{j_3+m_3+2j_1}\frac{1}{\sqrt{2j_3+1}}
C^{j_3m_3}_{j_1-m_1j_2-m_2}.
\]
The centrifugal potential is
\beq
v^{\lambda l}_{\alpha}=\frac{\hbar^2}{m} \Big[\frac{l(l+1)}{x^2}+
\frac{\lambda(\lambda+1)}{y^2} \Big],
\eeq
and nucleon-nucleon potentials are $v_{\alpha\alpha'}(x)=<\alpha|v({\bf
x})|\alpha'>=\delta_{\lambda\lambda'}
\delta_{ss'}\delta_{\sigma\sigma'}\delta_{JJ'}v^{\sigma J}_{ll'},$
where $v^{\sigma J}_{ll'}$ are the potential representatives in the two-body basis ${\cal Y}^{JJ_z}_{l\sigma}
({\bf\hat{x}})$ (most often abbreviated as $^{2\sigma+1}l_J)$.

    The set of partial differential equation Eqs.(\ref{fexy}) must be solved for functions satisfying the regularity conditions
\begin{equation}
\label{rcond}
\Phi^{\lambda_0 s_0 M_0}_{\alpha}(0,\theta)=\Phi^{\lambda_0 s_0 M_0}_{\alpha}(\rho,0)
=\Phi^{\lambda_0 s_0 M_0}_{\alpha}(\rho,\pi/2)=0
\end{equation}

    The asymptotic conditions for nd breakup scattering has the following form \cite{MerkAs}:
$$
\begin{array}{c}
\Phi^{\lambda_0s_0M_0}_{\alpha}(x,y)\sim
\Big\{\Big[\delta_{\lambda\lambda_0}\delta_{ss_0}\delta_{\sigma 1}
\delta_{j1}\hat{j}_\lambda(qy) + \Big(-\hat{y}_\lambda(qy)+i\hat{j}_\lambda(qy)\Big)
a^{M_0}_{\lambda s\lambda_0 s_0}\Big]\psi_l(x)
\\
 +O(y^{-1}) \Big \}+ A^{M_0}_{\alpha,\lambda_0 s_0}(\theta)\frac{e^{iKX}}{\sqrt{X}} +O(X^{-3/2}),
\end{array}
$$
\begin{equation}
\label{asympt}
\end{equation}
$$
X^2=x^2+y^2, \ \ K^2=\frac{mE}{\hbar^2}, \ \ x\ {\rm finite},\ \ y\to\infty,
$$
where  $\psi_l$ is $l-th$ component of deuteron wave function ($l=$0,2), and $\hat j$ and $\hat y$ are the regularized spherical Bessel functions.

The matrix of partial elastic amplitudes $a$ has the structure
\begin{equation}
a^{M_0}_{\lambda s\lambda_0 s_0} =\frac{\eta \exp (2i\delta)-1}{2i},
\end{equation}
where $\eta$ and $\delta$ are the inelasticities and scattering phases. In Eq.(\ref{asympt}) the amplitudes $ A^{M_0}_{\alpha,\lambda_0 s_0}$ are related with physical breakup amplitudes.
The inelastic scattering amplitudes are given by the sum
\beq
 {\cal
A}^{M_0}_{\alpha,\lambda_0s_0}(\theta)=
A^{M_0}_{\alpha,\lambda_0s_0}(\theta)+ \int_{-1}^1du\sum_{\beta}
g_{\alpha\beta}(\theta,u)A^{M_0}_{\beta,\lambda_0s_0}(\theta').
\label{bampl}
\eeq

     To simplify the numerical solution of Eqs.(\ref{fexy}), we write down Eqs.(\ref{fexy}) in the polar coordinate
system ($\rho^{2} = x^{2}+y^{2}$ and $\tan\theta = y/x$):
\begin{equation}
\label{fe}
\begin{array}{c}
\displaystyle
\Big[E+\frac{\hbar^2}{m}(\frac{\partial^2}{\partial\rho^2}
+\frac1{\rho^2}\frac{\partial^2}{\partial\theta^2}+\frac1{4\rho^2})
-v_{\alpha}^{\lambda l}(\rho,\theta)
\Big]U^{\lambda_0 s_0 M_0}_{\alpha}(\rho,\theta) =
\\
\\
\displaystyle
\sum_{\beta}v_{\alpha\beta}(\rho,\theta)\Big[U^{\lambda_0 s_0 M_0}_{\beta}(\rho,\theta)+
\int_{-1}^1du\sum_{\gamma}g_{\beta\gamma}(\theta,u,\theta'(\theta,u))
U^{\lambda_0 s_0 M_0}_{\gamma}(\rho,\theta')\Big].
\end{array}
\end{equation}
Here the first derivative in the radius is eliminated by the substitution $U = \rho^{-1/2}\Phi$.
In Eqs.(15-16) the angular variable $\theta^{\prime}$ is defined by
\begin{equation}
\cos ^2\theta ^{^{\prime }}(u,{\theta })=\frac 14\cos ^2\theta - \frac{\sqrt{%
3}}2\cos \theta \sin \theta \cdot u+\frac 34\sin ^2\theta.
\end{equation}
\subsection {Numerov method}
   Modification of the Numerov method for the set of the differential
equations (\ref{fe}) does not present any difficulties in principle.
As is well known, the Numerov method is an efficient algorithm for solving second-order differential equations. The important feature of the equations for the application of Numerov's method is that 
the first derivative has to be absent. The aim of this method is to improve the accuracy of the finite-difference approximation for the second derivative. Starting from the Taylor expansion truncated 
after the sixth derivative for two points adjacent to $x_n$, that is for $x_{n-1}$ and $x_{n+1}$ one sums these two expansions to give a new computational
formula that includes the fourth derivative. This derivative can be found by straightforward differentiation of the second derivative
from the initial second-order differential equation (see the details in \cite{Sus}). For brevity, we omit the corresponding derivation and present only the final formula of
Numerov's method for Eqs. (\ref{fe}), not indicating indices $\lambda_0s_0M_0 $:
$$
\begin{array} {l}
\displaystyle
-\Big[E + \frac{12}{(\Delta\rho)^2}+(1+\frac{2\Delta\rho}{\rho_j})\frac{T_{\alpha}(\theta)}{\rho^2_j}\Big]
U_{\alpha}(\rho_{j-1},\theta) =
\\
\displaystyle
\sum_{\beta}(v_{\alpha\beta}(\rho_j,\theta)-\Delta\rho v'_{\alpha\beta}(\rho_j,\theta))(U_{\beta}(\rho_{j-1},\theta)
+\sum_{\gamma}\int_{\theta^-}^{\theta^+}d\theta^{\prime}g_{\beta\gamma}(\theta,\theta')
U_{\gamma}(\rho_{j-1},\theta^{\prime}))
\\
\displaystyle
-2\Big[5E - \frac{12}{(\Delta\rho)^2}+(5+\frac{3\Delta\rho}{\rho_j})\frac{T_{\alpha}(\theta)}{\rho^2_j}\Big]
U_{\alpha}(\rho_j,\theta)
\\
\displaystyle
+\sum_{\beta}(10v_{\alpha\beta}(\rho_j,\theta)+(\Delta\rho)^2v^{\prime\prime}_{\alpha\beta}(\rho_j,\theta))
(U_{\beta}(\rho_j,\theta)+\sum_{\gamma}\int_{\theta^-}^{\theta^+}d\theta^{\prime}g_{\beta\gamma}(\theta,\theta')
U_{\gamma}(\rho_j,\theta^{\prime}))
\\
\displaystyle
-\Big[E+ \frac{12}{(\Delta\rho)^2}+(1-\frac{2\Delta\rho}{\rho_j})\frac{T_{\alpha}(\theta)}{\rho^2_j}\Big]
U_{\alpha}(\rho_{j+1},\theta)
\\
\displaystyle
+\sum_{\beta}(v_{\alpha\beta}(\rho_j,\theta)+\Delta\rho v'_{\alpha\beta}(\rho_j,\theta))(U_{\beta}(\rho_{j+1},\theta)
+\sum_{\gamma}\int_{\theta^-}^{\theta^+}d\theta^{\prime}g_{\beta\gamma}(\theta,\theta')
U_{\gamma}(\rho_{j+1},\theta^{\prime})) = 0,
\end{array}
$$
\begin{equation}
\label{reseq}
\end{equation}
where
$$T_{\alpha}(\theta)=\frac{\partial^2}{\partial\theta^2}-\frac{l(l+1)}{\cos^2\theta}
-\frac{\lambda(\lambda+1)}{\sin^2\theta}+\frac14.$$
In Eq. (\ref{reseq}) $\rho_j$ is the $j-th$ current point for hyperradius $\rho \in (0,R_{max})$ in the radial
grid ($j=1,2,\dots , N_{\rho}$), $\Delta\rho_j$ is the radial step-interval.

     To ensure the accuracy of order $(\Delta \theta)^4$ for the approximation in the angular variable, Hermitian splines of the fifth degree have been used (see Ref. \cite{KviH}). These splines are local 
and each spline $S_{\sigma i}(x)$ is defined for $x$ belonging to two adjacent subintervals $[x_{i-1},x_i]$ and $[x_i,x_{i+1}]$.
Their analytical form is fixed by the following smoothness conditions:
\begin{equation}
S_{\sigma i}(x_{i-1})=0,\ \ S_{\sigma i}(x_{i+1})=0, \ \ \sigma=0,1,2,
\end{equation}
and
\begin{equation}
\begin{array}{l}
S_{0i}(x_i)=1, \ \ \ \ S^{\prime}_{0i}(x_i)=0, \ \ \ \ S^{\prime\prime}_{0i}(x_i)=0,\\
S_{1i}(x_i)=0, \ \ \ \ S^{\prime}_{1i}(x_i)=1, \ \ \ \ S^{\prime\prime}_{1i}(x_i)=0,\\
S_{2i}(x_i)=0, \ \ \ \ S^{\prime}_{2i}(x_i)=0, \ \ \ \ S^{\prime\prime}_{2i}(x_i)=1.\\
\end{array}
\end{equation}
Expansion of the Faddeev component into basis of the Hermitian splines has the following form:
\begin{equation}
\label{sex}
U_{\alpha}(\rho,\theta)=\sum_{\sigma=0}^{2}\sum_{j=0}^{N_{\theta}+1}S_{\sigma j}(\theta)
C^{\sigma}_{\alpha j}(\rho),
\end{equation}
where $N_{\theta}+1$ is the number of internal subintervals for the angular variable $\theta \in [0,\pi/2] $.

     To reduce the resulting equation (\ref{reseq}) to an algebraic problem, one should explicitly calculate the derivatives of $NN$ potentials $v_{\alpha\beta}(\rho,\theta)$ with respect to $\rho$ and the second derivates
of splines $S_{\sigma j}(\theta)$ with respect to $\theta$. It is convenient to express the second derivative of
component $U_{\alpha}$ with respect to $\theta$ through $U_{\alpha}$ itself using Eq.(\ref{sex}). Upon substituting
the spline expansion (\ref{sex}) and expression for its second derivative into Eqs.(\ref{reseq}), we use a
collocation procedure with three Gaussian quadrature points per subinterval. As the number of internal breakpoints
for angular variable $\theta$ is equal to $N_{\theta}$, the basis of quintic splines consists of $3N_{\theta}+6$
functions. Three of them should be excluded using the last two regularity conditions from (\ref{rcond}) and
continuity of the first derivative in $\theta$ of the Faddeev component at either $\theta = 0$ or $\theta = \pi/2$,
as the collocation procedure yields $3N_{\theta}+3$ equations. Finally Eqs.(\ref{reseq}) for the Faddeev components
are to be written as the following matrix equation:
\begin{equation}
\label{algpr}
\begin{array}{l}
 \ \ \ \ \ \ \ \ \ \ \ \ \ \ \ \ \ \ A_{1}U_{1}+G_{1}U_{2}\ \  =0,\\
B_{j}U_{j-1}\ \ \ \ +A_{j}U_{j}+G_{j}U_{j+1}=0,\ \ \ \ \ \ \ \ \ \ \ \ \ \ j=2,...N_{\rho}-1,\\
B_{N_{\rho}}U_{N_{\rho}-1}+A_{N_{\rho}}U_{N_{\rho}}\ \ \ \ \ \ \ \ \ \ =-G_{N_{\rho}}U_{N_{\rho}+1}.
\end{array}
\end{equation}
In this equation vector $U_{k} = U(\rho_{k})$ has dimension $N_{in}$ and matrices  $A,B,G$, derived from Eq.(18), have dimension
$N_{in} \times N_{in}$ where $N_{in}$ = $N_{\alpha} \times N_c$, and $N_{\alpha}$ is the number of partial
waves and $N_c=3N_{\theta}+3$ is the number of collocation points in the angular variable $\theta$.
\subsection {The novel method of solution}
To derive equations for calculation of breakup nd amplitudes, the method of partial inversion \cite{Sus} has been applied. We write down Eq.(\ref{algpr}) in a matrix form:
\begin{equation}
(D*U)_i=-\delta_{iN_{\rho}}G_{N_{\rho}}U_{N_{\rho}+1}.
\end{equation}
Here matrix D is of dimension $N_{\rho}N_{in} \times N_{\rho}N_{in}$,
and $N_{\rho}$ is the number of breakpoints in the hyperradius  $\rho$. The form of this equation results from keeping the incoming wave in the asymptotic conditions (\ref{asympt}). As a consequence, 
the right hand part of Eq. (\ref{algpr}) has a single nonzero term marked with index $N_{\rho}+1$. Sparse (tri-block-diagonal) structure of matrix D optimizes considerably the inversion problem. Hyperradius $\rho_{n+1}=R_{max}$, where
$R_{max}$ is the cutoff radius at which the asymptotic conditions Eqs. (\ref{asympt}) are implemented. By formal inversion of the matrix D in Eq. (23), the solution of the problem may be written in the following form:
\begin{equation}
\label{solv}
  U_{j}=-D^{-1}_{jn}G_{N_{\rho}}U_{N_{\rho}+1},  \ \ \ \ j=1,2....N_{\rho}.
\end{equation}
 In Eq. (\ref{solv}) one should
consider the last two components of vector $U$:
\begin {equation}
\begin {array}{l}
U_{N_{\rho}-1}=-D^{-1}_{N_{\rho}-1N_{\rho}}G_{N_{\rho}}U_{N_{\rho}+1}\\
U_{N_{\rho}} \ \ \ =-D^{-1}_{N_{\rho}N_{\rho}}G_{N_{\rho}}U_{N_{\rho}+1}.
\end {array}
\end {equation}
Provided $R_{max}$ is large enough, the vectors $U_{N_{\rho}-1}, U_{N_{\rho}}$ on the left side
of Eqs.(25) may be replaced by the
corresponding vectors obtained by evaluating
Eqs. (\ref{asympt}) at the radii
 $\rho =\rho_{N_{\rho}-1}$ and $\rho =\rho_{N_{\rho}}$. As a result
we obtain a set of linear equations for the unknown amplitudes $a$ and $\cal A$:
\begin {equation}
\label{ipp}
\begin {array}{l}
 a\cdot \it v_{N_{\rho}-1} + \it m_{N{\rho}-1}\cdot{\cal A}={\cal F}_{N_{\rho}-1}\\
 a\cdot \it v_{N_{\rho}}\ \ \  + \it m_{N_{\rho}}\cdot{\cal A}\ \ \  ={\cal F}_{N{\rho}}.
\end {array}
\end {equation}
For the sake of brevity, we do not display here the explicit form of vectors
$\it v_{j}, {\cal F}_{j}$ and matrices $\it m_{j}$. As $R_{max} \rightarrow \infty$
the set of equations (\ref{ipp}) has a constant $a$ as a solution. At finite
$R_{max}$ its solution is a vector $a$ with generally different components corresponding to different angles. We follow the method of S.P. Merkuriev \cite{MerkAs}, which consists in selecting the components of $a$ 
in the region of the maximum of the deuteron wave function, where $a$ turns out to be independent of the angle.

     Furthermore, we propose a new method for a more adequate calculation of the amplitudes. The set of linear equations (\ref{ipp}) is over determined, since the number of equations is $2\cdot N_{in}$ and the number of unknowns is $N_{in}$+1. Therefore it is natural to use the least-squares method (LSM). One can apply it by two ways. In the first one, it is needed to express the breakup amplitude ${\cal A}$ from the  lower equation (\ref{ipp}) and substitute it into the upper one.
As a result one has the following expression:
\begin {equation}
a\cdot \bf v =\bf F,
\end {equation}
where vectors are defined as follows: $\bf v =\it v_{n-1}-\it m_{n-1}\it m^{-1}_{n}\it v_n $,
$\bf F = {\cal F}_{n-1}-\it m_{n-1}\it m^{-1}_{n}{\cal F}_n$.
According to LSM one should to minimize the following functional
\begin {equation}
\|a\cdot \bf v-\bf F\|^2=\min.
\end {equation}
Differentiating this expression in Re\,$a$ and Im\,$a$ we obtain
\begin {equation}
\label{afm}
  a = \frac {(\bf v^{\ast},\bf F)} {(\bf v^{\ast},\bf v)}\ \ ,
\end {equation}
where $(\xi^{\ast},f)$ is an ordinary scalar product.

  In the second way, it is needed to express the elastic amplitude $a$ from the lower equation (\ref{ipp}) using the scalar product:
\begin {equation}
\label{ama}
  a = \frac {(\it v^{\ast}_n,{\cal F}_n-\it m_n{\cal A})}
{(\it v^{\ast}_n,\it v_n)}.
\end {equation}
Substituting $a$ from Eq. (\ref{ama}) into the equation (\ref{ipp}),
leads to the set of linear equations
\begin {equation}
\label{at}
\it m_{n-1}{\cal A}-\it v_{n-1}\frac {(\it v^{\ast}_n,\it m_n {\cal A})}
{(\it v^{\ast}_{n},\it v_n)}={\cal F}_{n-1}-\it v_{n-1}\frac{(\it v^{\ast}_n,{\cal F}_n)}
{(\it v^{\ast}_n,\it v_n)}.
\end {equation}
The explicit form of Eq. (\ref{at}) is as follows
\begin{equation}
\label{atm}
\begin{array}{l}
\sum^{N_c}_{j=1}\Big\{\it m_{n-1,ij}-\frac{\it v_{n-1,i}}{(\it v^{\ast}_{n},\it v_n)}
\sum^{N_c}_{k=1}\it v^{\ast}_{n,k}\it m_{n,kj}\Big \}{\cal A}_j =  
\\
\\
\qquad {\cal F}_{n-1,i}-\it v_{n-1,i}\frac{(\it v^{\ast}_n,{\cal F}_n)}{(\it v^{\ast}_n,\it v_n)}
,\ \ i=1,...,N_c.
\end{array}
\end{equation}
Solving the set in Eq. (\ref{atm}), we get the breakup amplitude $\cal A$. Substituting the obtained breakup amplitude into Eq. (\ref{ama}), one may compute the elastic amplitude $a$. Note that one can apply 
Eq. (\ref{ama}) to calculate the elastic amplitude $a$ either in the components or via a scalar product. In the first case, the components of $a$ are practically equal to a constant for all angles $\theta \in (0,\pi/2)$ 
and this constant coincides with the value of $a$ calculated by using the scalar product to the fourth decimal. It should also be noted that the elastic amplitudes calculated by the method from \cite{MerkAs} 
and LSM coincide with this constant to the same accuracy. To control the accuracy of calculations, all methods are used.
\subsection{Observables}
     To calculate observables for elastic scattering of nucleon from deuteron in the direction ${\bf\hat{q}'}$ (initial direction ${\bf\hat{q}}$ is along the z-axis), one has to derive the equation for the elastic 
amplitude as a function of scattering angle. Omitting this derivation, we represent the final expression for this amplitude in MGL basis:
\begin{equation}
\label{hata}
\begin{array}{c}
\displaystyle
\hat{a}_{\sigma'_z,J'_z,\sigma_z,J_z}({\bf\hat{q}'}) =
\sum_{M}\sum_{\lambda' s'}\sum_{\lambda s}i^{\lambda-\lambda'}\sqrt{\frac{2\lambda+1}{4\pi}}
\\
\displaystyle
C^{MM_z}_{\lambda' M_z-\sigma'_z-J_z',s' \sigma_z'+J_z'}
C^{MM_z}_{\lambda 0, s \sigma_z+J_z}C^{s' \sigma_z'+J_z'}_{1/2 \sigma_z', 1 J_z'}
C^{s \sigma_z+J_z}_{1/2 \sigma, 1 J_z}Y_{\lambda' M_z-\sigma_z'-J_z'}(
{\bf\hat{q'}})a^M_{\lambda' s' \lambda s},
\end{array}
\end{equation}
with $M_z=\sigma_z+J_z$.

     In Eq. (\ref{hata}) $\sigma' \sigma_z' (\sigma, \sigma_z)$ and $J' J_z'( JJ_z)$ are spin and its projection for incoming (scattered) nucleon, and the deuteron in the rest (scattered deuteron), respectively. 
Thus, the nuclear part of the elastic amplitude is a $(2\times 2)\otimes(3\times 3)$ matrix in the spin states of nucleon and deuteron, depending on the spherical angles $\theta$ and $\phi$.

     The spin elastic observable formulas can be taken from the review \cite{Gloec}. They are expressed via spin $2\times 2$ matrices $\sigma_i$ for the nucleon and $3\times 3$ matrices ${\cal P}_i$ and 
${\cal P}_{ik}$ for the deuteron. The latter are related to the deuteron spin matrices $S_i$:
\begin{equation}
S_x=\frac{1}{\sqrt{2}}\left ( \begin{array}{ccc}
    0     & 1 & 0\\
    1 & 0 & 1\\
    0& 1& 0
    \end{array} \right),\
S_y=\frac{1}{\sqrt{2}}\left ( \begin{array}{ccc}
    0     & -i & 0\\
    i & 0 & -i\\
    0& i& 0
    \end{array} \right),\
S_z=\left ( \begin{array}{ccc}
    1     & 0 & 0\\
    0 & 0 & 0\\
    0& 0& -1
    \end{array} \right).\
\end{equation}
One has ${\cal P}_i=S_i$, ${\cal P}_{ik}=3/2(S_iS_k+ S_kS_i)$, ${\cal P}_{zz}=3S_zS_z-2I$, and
${\cal P}_{xx}-{\cal P}_{yy}=3(S_xS_x- S_yS_y)$.

Nucleon analyzing powers $A_k$ are
 \beq
 A_k=\frac{{\rm
Tr}\,(\hat{a}\sigma_k\hat{a}^{\dagger})} {{\rm Tr}\,(\hat{a}\hat{a}^{\dagger})}.
\eeq
If the scattering plane is the $xy$ plane and the $y$ axis points to the direction $\bf{q}\times\bf{q'}$ then due to parity conservation $A_x=A_z=0$ and the only non-zero component is $A_y$.
The deuteron vector and tensor analyzing powers are defined as
\beq
A_k=\frac{{\rm Tr}\,(\hat{a}{\cal P}_k\hat{a}^{\dagger})} {{\rm
Tr}\,(\hat{a}\hat{a}^{\dagger})},\ A_{jk}=\frac{{\rm
Tr}\,(\hat{a}{\cal P}_{jk}\hat{a}^{\dagger})} {{\rm
Tr}\,(\hat{a}\hat{a}^{\dagger})}.
 \eeq
Parity conservation puts $A_x,A_z,A_{xy}$ and $A_{yz}$ to zero. So the non-vanishing and independent analyzing powers are defined by
\beq
iT_{11}=\frac{\sqrt{3}}{2}A_y,\ \ T_{20}=\frac{1}{\sqrt{2}}A_{zz},\
\ T_{21}=-\frac{1}{\sqrt{3}}A_{xz},\ \
T_{22}=\frac{1}{2\sqrt{3}}(A_{xx}-A_{yy}).
\eeq

Also spin transfer coefficients are given in the review  \cite{Gloec}. They have the same structure as the quantities above, with slightly different matrices to be inserted between $\hat{a}$ and
$\hat{a}^{\dagger}$.

In the case of Nd breakup scattering expression for physical breakup amplitude to calculate breakup observables has much more complex form. Below we present the main details of its derivation.
The asymptotic of the wave function with the given incident plane wave is defined as follows
\beq
\begin{array}{c}
\displaystyle
\Psi_{q,s_0,s_{0z}}({\bf x,y})
= \\
\displaystyle
\frac{e^{iKX}}{X^{5/2}}\frac{4\pi}{q}\sum_{\lambda_0,\lambda_{0z},M_0,M_{0z}}
c^{\lambda_0,s_0,M_0}_{\lambda_{0z},s_{0z}}({\bf\hat{q}})
\sum_{\alpha} \frac{{\cal A}^{\lambda_0 s_0,M_0}_\alpha(\theta)}{\sin\theta\cos\theta}Z_{\alpha}({\bf \hat{y},\hat{x}}),
\end{array}
\label{psi11}
\eeq
where the initial plane wave has the form
\beq c^{\lambda_0 s_0 M_0}_{\lambda_{0z}s_{0z}}({\bf
\hat{q}})= i^{\lambda_0}
C^{M_0M_{0z}}_{\lambda_0\lambda_{0z}s_0s_{0z}}
[Y_{\lambda_0\lambda_{0z}}]^*({\bf \hat{q}}),
\label{ccoef}
\eeq
and $Z_{\alpha}({\bf \hat{x},\hat{y}})$  are the spin-angular-isospin eigenfunctions.

     The exponential factor is multiplied by the function depending only on angles, that are directions of vectors ${\bf q}$, ${\bf x}$ and ${\bf y}$. This gives a probability to find 
the breakup particles at given directions and so with the amplitude for the breakup with given directions of ${\bf p}'$ and ${\bf q}'$. So the breakup amplitude is
\beq
{\cal A}_{q,s_0,s_{0z}}({\bf p}', {\bf q}')=
\frac{4\pi}{q}\sum_{\lambda_0,\lambda_{0z},M_0,M_{0z}}
c^{\lambda_0,s_0,M_0}_{\lambda_{0z},s_{0z}}({\bf\hat{q}})
\sum_{\alpha} \frac{{\cal A}^{\lambda_0 s_0,M_0}_\alpha(\theta)}{\sin\theta\cos\theta}
Z_{\alpha}({\bf \hat{q}',\hat{p}'}),
\label{ain}
\eeq
where now $\theta=\arctan (q'/p')$

The only thing necessary to pass to the formula for the breakup is to project this amplitude onto the state with given projections of spins of the three particles $\mu_1,\mu_2,\mu_3$. 
This will obviously be given by introducing into the sum over $\alpha$ in Eq. (\ref{ain}) the projection
\beq
d_{\alpha,\mu_1\mu_2\mu_3}^{M_0M_{0z}}({\bf {\hat p}'{\hat ,q}'})
=<\mu_1\mu_2\mu_3|Z_{\alpha}({\bf \hat{q}',\hat{p}'})>.
\label{dcoef}
\eeq

As the result we get the breakup scattering amplitude as a function of final nucleon momenta in the following form
\[
{\cal A}(\hpp,\hqp,\mu_1\tau_1\mu_2\tau_2\mu_3\tau_3|{\bf q},
s_0,s_{0z})\]\beq=
\frac{4\pi}{q}\sum_{\alpha,\pi,\lambda_0,M_0,\lambda_{0z},M_{0z}}
d^{M_0,M_{0z}}_{\mu_1\tau_1\mu_2\tau_2\mu_3\tau_3,\alpha}
 (\hpp,\hqp)
\frac{{\cal A}^{M_0}_{\alpha,\lambda_0,s_0}(\theta')}
{\sin{\theta'}\cos{\theta'}}
c^{\lambda_0s_0M_0}_{\lambda_{0z},s_{0z}}(\hqp)
\eeq
where $\mu_i$ and $\tau_i$ $i=1,2,3$ are the spin and isospin projections of the three nucleons, $\theta'=\arctan(q'/p')$, and summation goes over $\lambda_{0z}+s_{0z}=M_{0z}$.
${\cal A}^{M_0}_{\alpha,\lambda_0,s_0}$ is the spherical inelastic
amplitude defined in Eq. (\ref{bampl}). The $d$-coefficients are
\[
d^{M_0,M_{0z}}_{\mu_1\tau_1\mu_2\tau_2\mu_3\tau_3,\alpha} (
\hpp,\hqp)=(-1)^{\lambda+J+M-1/2}[(2J+1)(2s+1)]^{1/2}C^{\sigma \mu_2+\mu_3}_{\frac{1}{2} \mu_2,\frac{1}{2} \mu_3}\]\[
C^{t \tau_2+\tau_3}_{\frac{1}{2} \tau_2,\frac{1}{2} \tau_3}
C^{\frac{1}{2} -\frac{1}{2}}_{\frac{1}{2} -\frac{1}{2}-\tau_2-\tau_3,t \tau_2+\tau_3}
\sum_{LS}[(2L+1)(2S+1)]^{1/2}C^{ SS_z}_{\frac{1}{2} \mu_1,\sigma \mu_2+\mu_3}
C^{M_0M_{0z}}_{LM_{0z}-S_z,SS_z}\]
\beq
 \left \{ \begin{array}{ccc}
    l     & \sigma & J \\
    1/2 & s & S
    \end{array} \right \}
 \left \{ \begin{array}{ccc}
    \lambda     & l & L \\
    S & M & s
    \end{array} \right \} {\cal Y}^{L,M_{0z}-S_z}_{\lambda l}(\hqp,\hpp)
\eeq
 where $S_z=\mu_1+\mu_2+\mu_3$ and $t$=0 or 1
 according to antisymmetry condition $l+\sigma+t$ odd.

The breakup differential cross-section is then
\beq
\frac{d^5\sigma}{d^3p'd^2\hqp}= \frac{4q'}{3K^3q} |{\cal
A}(\hpp,\hqp,\mu_1\tau_1\mu_2\tau_2\mu_3\tau_3|{\bf q},
s_0,s_{0z})|^2
\eeq
Note that
$d^3p'q'd^2\hat{q}'=p'd^2\hat{p}'d^3q'$

This formula may be transformed to the form which is used by the
experimentalists in the lab. system.
\beq
\frac{d^5\sigma}{dSd^2\hat{\bf k}_1d^2\hat{\bf k}_2}=
\frac{\sqrt{3}mk_1^2k_2^2}{qK^3\sqrt{D}} | {\cal A}(\hpp,\hqp,\mu_1\tau_1\mu_2\tau_2\mu_3\tau_3|{\bf q},
s_0,s_{0z})|^2
\label{d}
\eeq
where
\beq
D=k_1^2\Big(2 k_2-\hat{\bf k}_2({\bf k}_{lab}-{\bf k}_1)\Big)^2+
 k_2^2\Big(2k_1-\hat{\bf k}_1({\bf k}_{lab}-{\bf k}_2)\Big)^2
\label{Dk}
\eeq
and $S$ is the arc lenght along the allowed curve in the $E_1-E_2$ plane:
\beq
dS=dE_1\sqrt{1+\left(\frac{k_2}{k_1} \frac{ 2k_1 -\hat{\bf k}_1({\bf
k}_{lab}-{\bf  k}_2) }{ 2k_2 -\hat{\bf k}_2({\bf k}_{lab}-{\bf
k}_1)}\right)^2}
\eeq

This cross-section refers to the breakup experiment in which all individual spins and isospins of the final nucleons and the spin of the initial neutron-deuteron system are given. 
The unpolarized cross-section with the isospin projections of the final nucleons given is obtained by summing Eq. (\ref{d}) over $\mu_1, \mu_2, \mu_3$ and averaging over $s_0$ and 
its projections $s_{0z}$:
\beq
\begin{array}{c}
\displaystyle
\frac{d^5\sigma}{dSd^2\hat{\bf k}_1d^2\hat{\bf k}_2}= \\
\displaystyle
\frac{\sqrt{3}mk_1^2k_2^2}{qK^3\sqrt{D}}\frac{1}{6}  
 \sum_{\mu_1,\mu_2,\mu_3,s_0,s_{0z}} | {\cal A}(\hpp,\hqp,\mu_1\tau_1\mu_2\tau_2\mu_3\tau_3|{\bf q},
s_0,s_{0z})|^2
\end{array}
\eeq

To write this expression in a more convenient form, one can introduce a matrix between initial and final spins:
\beq
\hat{\cal A}(\hpp,\hqp|{\bf q})_{\mu_1,\mu_2,\mu_3|s_0,s_{0z}} \equiv
{\cal A}(\hpp,\hqp,\mu_1\tau_1\mu_2\tau_2\mu_3\tau_3|{\bf q},
s_0,s_{0z})
\eeq

In terms of this matrix the unpolarised cross-section can be written as
\beq
\frac{d^5\sigma}{dSd^2\hat{\bf k}_1d^2\hat{\bf k}_2}=
\frac{\sqrt{3}mk_1^2k_2^2}{qK^3\sqrt{D}}\frac{1}{6}{\rm Tr}\{\hat{\cal A}^{\dagger} \hat{\cal A}\}
\eeq

Using this matrix one can write all polarization observables in the same form as for the elastic channel. In particular the final proton analasing power will be given by the formula
\beq
A_k=\frac{{\rm Tr}\{\hat{\cal A}^{\dagger}\sigma^{(1)}_k \hat{\cal A}\}}{{\rm Tr}\{\hat{\cal A}^{\dagger} \hat{\cal A}\}},
\eeq
where $\sigma_{\mu_1^{\prime}\mu_1}$ is the spin matrix of the 1st nucleon, supposed to be proton $(\tau_1 = +1/2)$.
\section{Results}

\begin{figure}
\includegraphics[width=120.0mm]{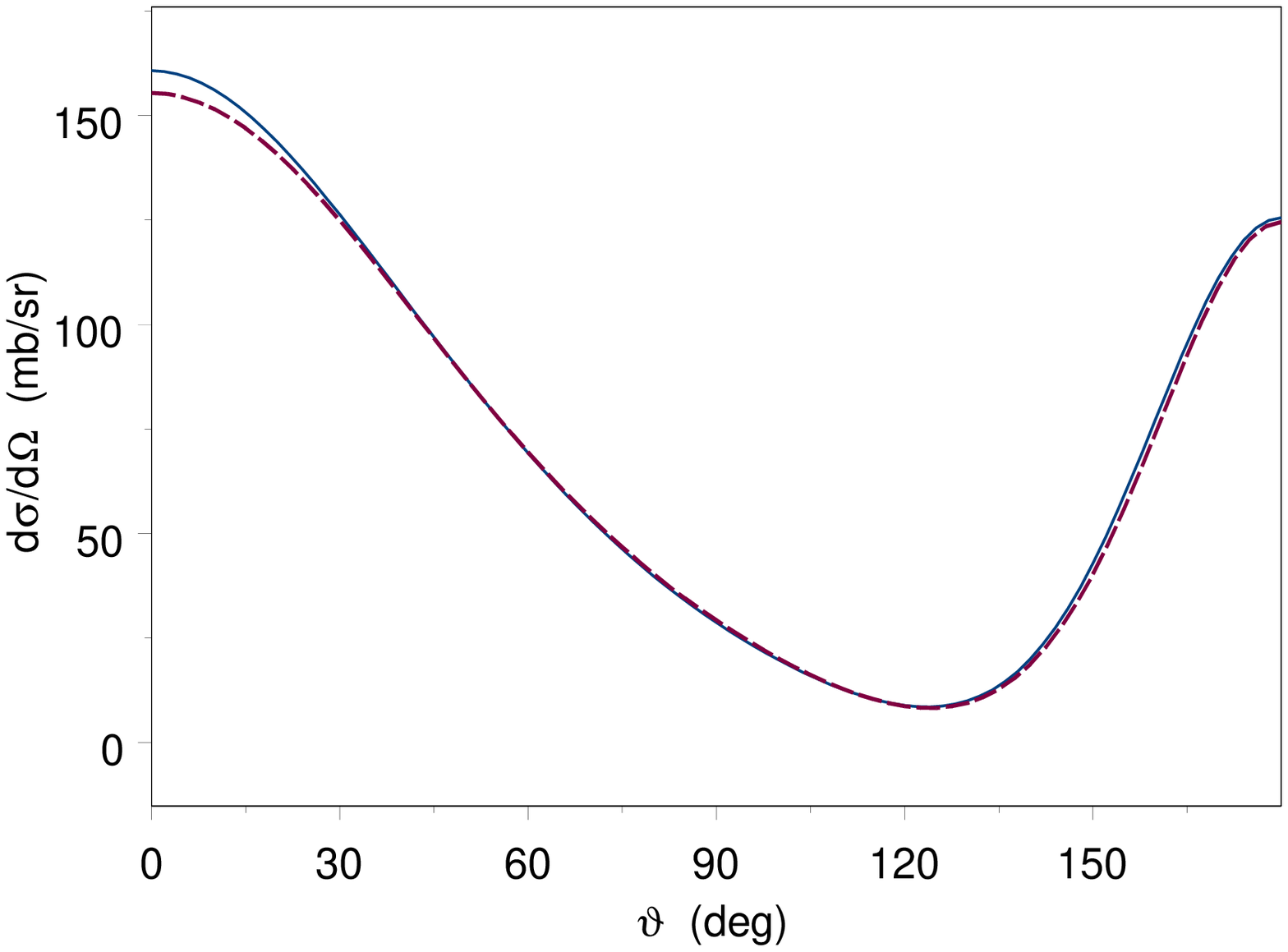}
\caption{\label{fig:1}
The differential cross section for $nd$ scattering at $\rm E_{lab}$=14.1~MeV. The solid line is our results. The dashed line corresponds to prediction of
 the Bochum group Ref. \cite{Wit}.
}
\end{figure}

\begin{figure}
\includegraphics[width=120.0mm]{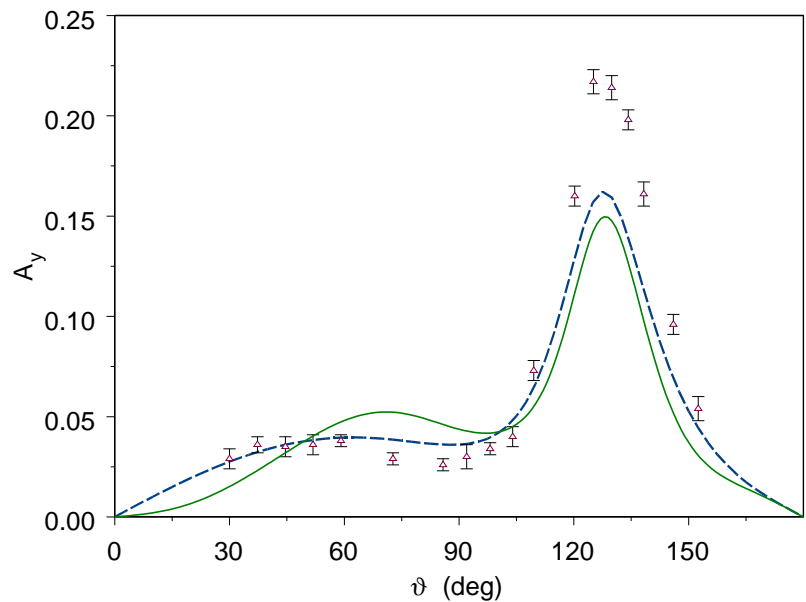}
\caption{\label{fig:2}
The nucleon analyzing power $\rm A_y$ for $nd$ scattering at $\rm E_{lab}$ =14.1~MeV. The solid line is our results. The dashed one is results of 
the Bochum group Ref. \cite{Wit}. 
The experimental data are from Ref. \cite{How}.
}
\end{figure}

\begin{figure}
\includegraphics[width=120.0mm]{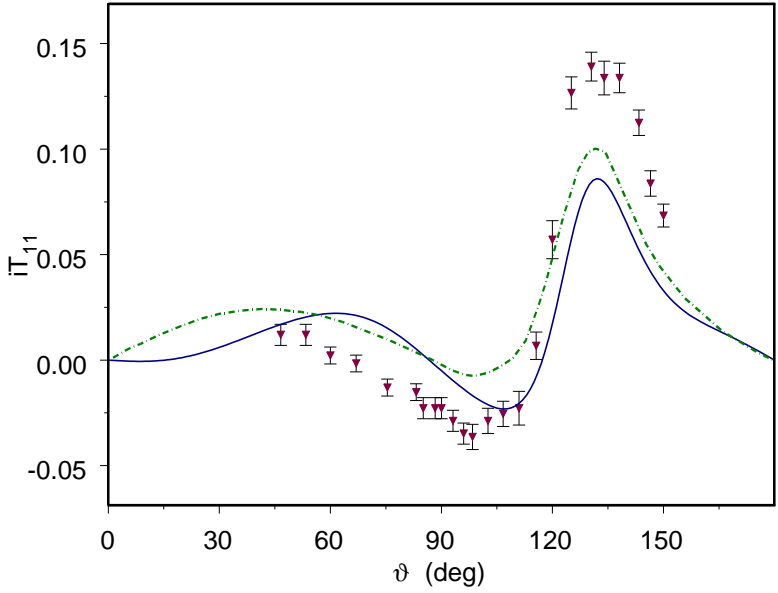}
\caption{\label{fig:3}
The deuteron analyzing power i$\rm T_{11}$  for $nd$ scattering at $\rm E_{lab}$ =14.1~MeV. The dot-dashed line is prediction of 
the Grenoble Group with SSC NN interaction Ref. \cite{Gign}. The proton-deuteron data at  $\rm E_{lab}$ =15.0 MeV are from Ref. \cite{McK}.
}
\end{figure}

We present our results for the differential cross section, nucleon
analyzing power $\rm A_y$, and deuteron vector $i\rm T_{11}$ for
$nd$ elastic scattering at 14.1 MeV in Figs. \ref{fig:1}--\ref{fig:3}
In all our calculations we used the AV14 NN potential. The total
angular momentum of the pair of nucleons $j_{23}$ has been taken
up to 3 and all values of the total three-nucleon angular momentum
$\rm M$ up to 13/2 with both parity values ($\pm$1) have been
taken into account (up to 62 partial waves in all).

 Fig. \ref{fig:1} shows our results for elastic differential
cross sections along with the prediction of the Bochum group Ref.
\cite{Wit} using the charge-dependent AV18 NN potential.  In Fig.
\ref{fig:2} our results for nucleon analyzing power are shown
together with the prediction of the Bochum group and experimental
data from \cite{How}. In general the agreement between our
calculations of $\rm A_y$ and those of the Bochum group is
satisfactory, since using different NN potentials may explain
small differences between these  calculations around the maximum
values of $\rm A_y$ (Fig. \ref{fig:2})

Our predictions for the deuteron analyzing power $i\rm T_{11}$ and
those of the Grenoble group \cite{Gign} are presented in Fig.
\ref{fig:3} along with the experimental data \cite{McK}. Again
there is a small disagreement between two calculations which can
be explained by the use of two different type of NN potentials.
The results of the Grenoble group have been obtained with NN
interactions of de Tourreil and Sprung, effective in the states
$^1S_0, ^3S_1, ^3D_1, ^1P_1, $ and $ ^3P_{0,1,2}$.

\begin{figure}
\includegraphics[width=120.0mm]{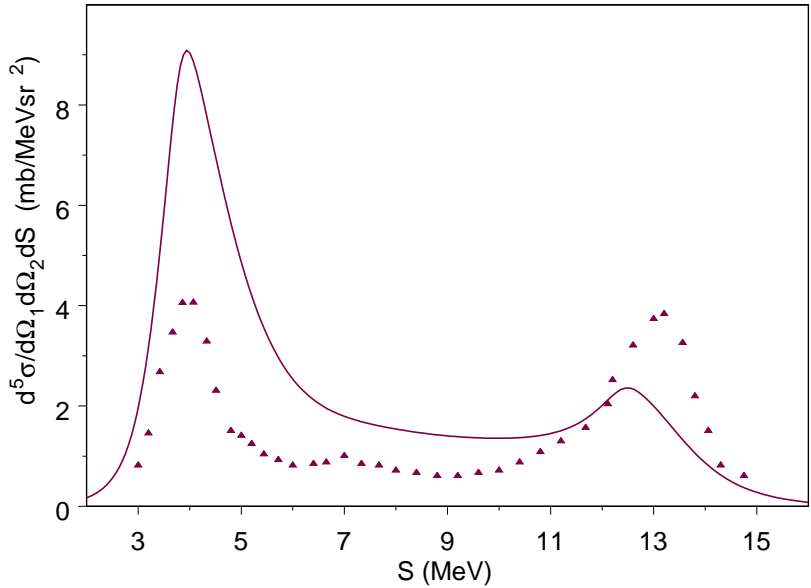}
\caption{\label{fig:4}
$nd$ breakup differential cross section as a function of the arc length $S$ under FSI configuration ($\theta_1$ =$52.6\,^{\circ}$, $\theta_2$ = $40.5\,^{\circ}$,
 $\phi_{12}$ =$180\,^{\circ}$) at  $\rm E_{lab}$ =14.1 MeV. The proton-deuteron experimental data denoted by solid triangles are from Ref. \cite{Karu}.
}
\end{figure}

\begin{figure}
\includegraphics[width=120.0mm]{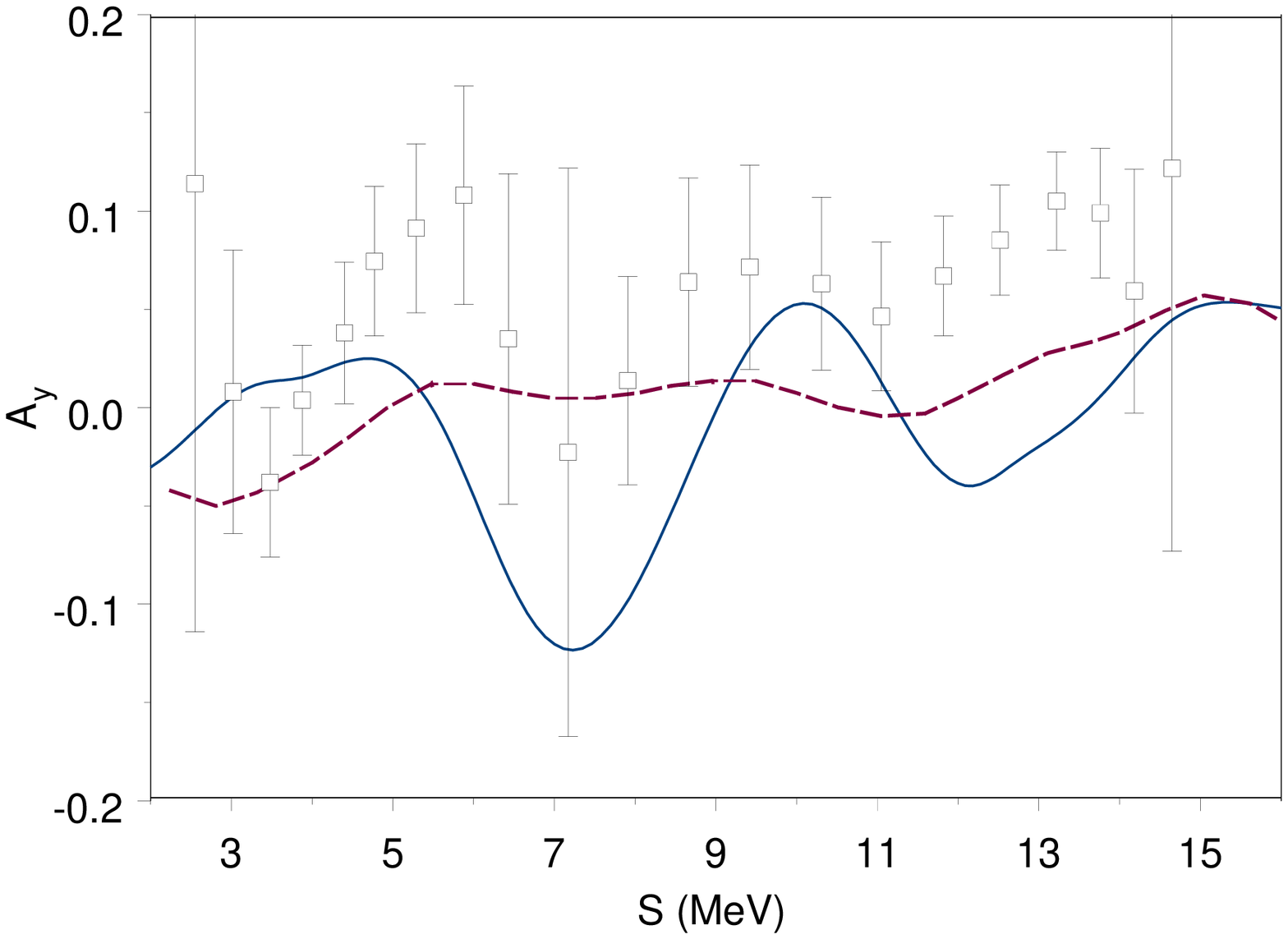}
\caption{\label{fig:5}
 The nucleon analyzing power as a function of the arc length $S$ under FSI configuration ($\theta_1$ =$52.6\,^{\circ}$, $\theta_2$ = $40.5\,^{\circ}$,
$\phi_{12}$ =$180\,^{\circ}$) at $\rm E_{lab}$ =14.1 MeV. The solid line is our results. The dashed one is results of Ref. \cite{WitG}.
The proton-deuteron experimental data denoted by open squares are from Ref. \cite{Karu}.
}
\end{figure}

For  nd breakup scattering at 14.1 MeV, our preliminary results
for the breakup differential cross section and nucleon analyzing
power $\rm A_y$ under FSI configuration are shown in Figs.
\ref{fig:4},\ref{fig:5} along with the experimental data
\cite{Karu}. Results for $\rm A_y$ are presented together with the
calculations of Witala et al. \cite{WitG} using the Paris NN
potential and  Bonn potential of version A and B with total
two-nucleon angular momentum ${ j\leq2}$. Thus in their work up to
34 partial wave states have been taken into account for each total
angular momentum $J$ and parity states.

 \section {Discussion}
Our results for $nd$ elastic scattering at 14.1~MeV and those from
the Bochum and Grenoble group are in  fair agreement. Some
differences can be attributed  to smaller values of the total
two-nucleon angular momentum $j$ taken into account in their
calculation as well as to different NN potentials used. In the
energy region of tens~MeV \cite{How} and \cite{McK} theoretical
predictions are 25-30\% lower than the experimental data.

For $nd$ breakup observables our predictions compare only
qualitatively with the experimental data and  predictions for $\rm
A_y$ of the Bochum group. Still the positions of the peaks in the
breakup angular distribution are correct. For the nucleon
analyzing power  the deep minimum at arc length $S$ equal to 7 in
our results  disagrees with predictions for $\rm A_y$  from
\cite{WitG}.

 \section {Conclusion}
For $nd$ elastic observables at 14.1 MeV acceptable agreement (in
view of different NN potentials) between predictions of our
calculations and those of the Bochum and Grenoble group
demonstrates the soundness of our novel method providing thereby a
new approach for calculating three-nucleon scattering. Our
approach can and will be used to include the Coulomb force. It is
well-known that $Nd$ polarization observables are the magnifying
glass for studying $^3P_J$ NN states and calculations which
rigorously include nuclear and electromagnetic interactions are
very important.

Our next step is to study pd breakup scattering using a new and
more accurate version of the Faddeev equations in configuration
space. In presence of the Coulomb interaction serious changes are
to be  made in view of $p$ and $n$ being different particles, so
that use of the isotopic formalism becomes unreasonable. As a
result the Faddeev equations in configuration space derived by
S.P. Merkuriev {\it et al.} in \cite{KKM} become only approximate.
Probably the error of their calculations of the elastic $pd$
scattering amplitudes is not  significant. However in the case of
$pd$ breakup scattering one has to use both charge-dependent NN AV18
potential and the new correct Faddeev equations.

\ack
We are very grateful to Prof. H. Witala for courteously giving the
computer code to calculate $nd$ observables. This work was supported
by NSF CREST award HRD-0833184 and NASA award NNX09AV07A. The work
of I.S. was supported in part by the Croatian Ministry of Science.

%
\newpage
\newpage


\section*{References}
\begin{thebibliography}{27}
\bibitem{Gloec}  Gl\"ockle W,  Witala H,  H\"uber D,  Kamada H,  Golak J
1996 Phys. Rep. {\bf 274} 107
\bibitem{Delt1}  Deltuva A {\it et al.} 2003 Phys. Rev. C{\bf 68}
024005; Deltuva A {\it et al}. 2007 Nucl. Phys. A{\bf 790} 52
\bibitem{Wir1}  Wiringa R B,  Smith R A, and  Ainsworth T L 1984 Phys. Rev. C{\bf29} 1207
\bibitem{Wir2}  Wiringa R B {\it et al}. 1995 Phys. Rev. C{\bf 51} 38
\bibitem{Mach1}  Machleidt R{\it et al}. 1996 Phys. Rev. C{\bf 53} R1483
\bibitem{Stoc1}  Stocks V G J {\it et al}.1994 Phys. Rev. C{\bf49} 2950
\bibitem{Coon} S.A. Coon {\it et al}. 1979 Nucl. Phys. {\bf 317} 242;
Coon S A and Han H K 2001 Few-Body Syst.  {\bf 30} 131 
\bibitem{Pudi}  Pudimer B S {\it et al}. 1995 Phys. Rev. Let. {\bf 51} 4396;
Pieper S C {\it et al}. 2001 Phys. Rev. C{\bf64}, 014001
\bibitem{Mach2} Machleidt R 2007 Nucl. Phys. A{\bf 790} 17 and references therein; 
Meissner U-G 2007 Nucl. Phys. A{\bf 790} 129
\bibitem{Ivo} Slaus I 2007 Nucl. Phys. A{\bf 790} 199
\bibitem{Torn} Tornow W {\it et al}. 1982 Phys. Rev. Lett. {\bf 49} 312;
Gruebler W {\it et al}. 1983 Nucl. Phys. A{\bf 398} 445; 
Neidel E M {\it et al}. 2003 Phys. Lett. B{\bf552} 29 and
references therein.
\bibitem{Wiel} Wielinga B J {\it et al}. 1976 Nucl. Phys. A{\bf 261} 13;
 Setze H R {\it et al}. 1996 Phys. Lett. B{\bf 388} 229 and
references therein;  Zhou Z {\it et al}. 2001 Nucl. Phys. A{\bf 684} 545
\bibitem{Koeln} Ley J {\it et al}. 2006 Phys. Rev. C{\bf 73} 064001
\bibitem{Maeda} Maeda Y {\it et al}.2011 Journal of Physics: Conf. Series {\bf 312}  082031
\bibitem{CRH}  Couture A H {\it et al}. 2012 Phys. Rev. C{\bf 85} 054004
\bibitem{MGL}  Merkuriev S P,  Gignoux C and Laverne A 1976 Ann. Phys. {\bf 99},
30
\bibitem{KKM}  Kvitsinsky A A,  Kuperin Yu A,  Merkuriev S P,  Motovilov A K and   Yakovlev S L 1986  Elem. Chastis At. Yadra {\bf 17} 267
\bibitem{MerkAs}  Merkuriev S P 1980 Ann. Phys. (N.Y.) {\bf 130} 3975;\\
 Merkuriev S P 1981 Acta Physica (Austriaca) Suppl. {\bf XXIII} 65
\bibitem{Sus}  Suslov V M and  Vlahovic B 2004 Phys. Rev. C{\bf 69} 044003
\bibitem{KviH}  Kvitsinsky A A and  Hu C-Y 1992 Few-Body Syst. {\bf 12} 7;
1998 Phys. Rev. C{\bf 58} 3085
\bibitem{Wit} H. Witala, private communication.
\bibitem{How}  Howel C R {\it et al}. 1987 Few-Body Syst. {\bf 2} 19
\bibitem{Gign}  Benayoun J J,  Chauvin J,  Gignoux C, and  Laverne A 1976 Phys. Rev. Lett. {\bf 36} 1438
\bibitem{McK}  McKee J S C {\it et al}. 1972 Phys. Rev. Lett. {\bf 29} 1613
\bibitem{Karu} M. Karus {it\ et al}. 1985 Phys. Rev. C{\bf 31} 1112
\bibitem{WitG}  Witala H, Gl\"ockle W and  Cornelius Th. 1989 Phys. Rev. C{\bf 39} 384
\end{thebibliography}
\end{document}